\documentclass[onecolumn,showpacs,preprintnumbers,amsmath,amssymb]{revtex4}
\usepackage{color}
\usepackage{subfigure}
\usepackage{amsmath,epsfig}
\usepackage{graphicx}
\usepackage{epstopdf}
\usepackage{dcolumn}
\usepackage{bm}
\usepackage{multirow}
\usepackage{hhline}
\usepackage{afterpage}
\usepackage{hyperref} 
\usepackage{float}
\usepackage{booktabs}

\begin{document}


\title{Longitudinal flow decorrelation in heavy-ion collision at RHIC energies using a multi-phase transport model}
\author{Prabhupada Dixit and Md.~Nasim}

\begin{abstract}
We present a study on the longitudinal flow decorrelation in heavy-ion collisions at the RHIC Beam Energy Scan (BES) energies ($\sqrt{s_{NN}}$ = 11.5 to 200 GeV in Au+Au collisions) using the AMPT model. We measure the second and third order factorization ratios ($r_{2}$ and $r_{3}$) across BES energies, finding $r_{2}$ weakly dependent on collision energy while $r_{3}$ shows a more prominent collision energy dependent. The effect of parton-parton scattering cross section on $r_{2}$ and $r_{3}$ is studied which suggests that longitudinal decorrelation could be a potential observable to constrain transport properties of the medium. We also analyze the contributions of flow-plane and flow magnitude decorrelation, with flow-plane decorrelation being dominant. we measure a recently proposed observable, the four particle cumulant ($T_{2}$), which remains resilient to non-flow effects and exhibits sensitivity to different decorrelation patterns. Through the measurement of $T_{2}$, we consistently observe a hint of  S-shaped or torqued decorrelation across all energy ranges in peripheral collisions (40-80\%). But in central and mid-central collisions, presence of any specific pattern of decorrelation is missing in AMPT model.
\end{abstract}

\maketitle

\section{Introduction}
Relativistic heavy-ion collisions can produce an exotic state of matter known as the Quark-Gluon-Plasma (QGP), where quarks and gluons are no longer confined within hadrons~\cite{QGP, QGP-2}. Azimuthal anisotropic flow is an important observable that provides sensitivity to the initial state and transport properties of the QGP medium formed in these collisions~\cite{flow1, flow2, flow3, flow4, flow5, flow6, flow7, flow8}. Under the assumption of longitudinal boost invariance, one would expect similar initial geometry and dynamics in both the forward and backward rapidity directions, as observed in the center-of-mass rapidity region ($y_{\text{cm}}$). However, deviations from this picture can occur, leading to the breaking of boost invariance. One observed signature of this is the longitudinal decorrelation of azimuthal anisotropic flow which arises due to event-by-event fluctuation of initial energy density distribution along rapidity direction. If one consider the transverse plane at different rapidity slices then the distribution of energy density in each slice could be different from other. This fluctuation in energy density distribution will lead to the fluctuation of flow angle and flow magnitude in each rapidity window. This phenomenon has been extensively studied through various experiments and models in A+A collisions at $\sqrt{s_{NN}}$ = 200 GeV and 2.76 TeV~\cite{exp-1, exp-2, exp-3, exp-4, ref-model-1, ref-model-2, ref-model-3, ref-model-4, ref-model-5, ref-model-6, ref-model-7, ref-model-8, ref-model-9, ref-1, ref-2, ref-3}. The Relativistic Heavy Ion Collider (RHIC) has initiated a beam-energy-scan (BES) program to investigate how various measurements change with beam energy. In this context, the measurement of flow decorrelation is crucial for understanding the initial state in heavy-ion collisions. The Multi-Phase Transport Model (AMPT) has been widely used to interpret the physics underlying experimental observations and is considered one of the most successful models for describing flow-related measurements in heavy-ion collisions at RHIC and LHC energies. In this study, we utilized the AMPT model to investigate flow decorrelation across a range of beam energies, from 11.5 GeV to 200 GeV. These calculations can be compared with upcoming measurements from the RHIC BES (Phase II) program.

\par
The flow decorrelation is quantified by comparing the flow in two symmetrically placed pseudorapidity windows, one in the positive side and the other in the negative side with respect to $\eta = 0$. The CMS collaboration has introduced a quantitative measure of decorrelation, known as the factorization ratio, defined as follows~\cite{exp-1}:

\begin{flalign}
\label{eqn:1}
r_{n}(\eta_{a}, \eta_{b}) &= \frac{\langle V_{n}(-\eta_{a})V_{n}^{*}(\eta_{b}))\rangle}{\langle V_{n}(+\eta_{a})V_{n}^{*}(\eta_{b}))\rangle} \\
&= \frac{\langle v_{n}(-\eta_{a})v_{n}(\eta_{b}) \cos n(\Psi_{n}(-\eta_{a}) - \Psi_{n}(\eta_{b}))\rangle}{\langle v_{n}(\eta_{a})v_{n}(\eta_{b}) \cos n(\Psi_{n}(\eta_{a})- \Psi_{n}(\eta_{b}))\rangle}.
\end{flalign}

Where the angular bracket represents the average over the events. $V_{n} = v_{n}e^{in\psi_{n}}$, is the flow vector and $\eta_{a}$ represents the pseudorapidity window symmetrically placed around $\eta = 0$, while $\eta_{b}$ denotes the reference pseudorapidity window typically located in the forward or backward $\eta$ region. $v_{n}$ and $\Psi_{n}$ correspond to the $n$-th order flow magnitude and event plane angle, respectively. In the absence of longitudinal decorrelation, the value of $r_{n}(\eta_{a}, \eta_{b})$ would be unity. However, in the presence of decorrelation, the factorization breaks down, resulting in a ratio smaller than unity. This is due to a stronger correlation between $\eta_{a}$ and $\eta_{b}$ compared to the correlation between $-\eta_{a}$ and $\eta_{b}$.
\par
The breaking of the factorization ratio can be attributed to two potential effects: flow plane decorrelation and/or flow magnitude decorrelation. Previous studies~\cite{ref-2, ref-3} have separately investigated these two effects using hydrodynamical models. Flow magnitude decorrelation can be assessed by maintaining the same event plane angle for both $\eta_{a}$ and $-\eta_{a}$ in Eq.~\ref{eqn:1}, resulting in a simplified form:

\begin{equation}
\label{eqn:2}
r_{n}^{v} = \frac{\langle v_{n}(-\eta_{a})v_{n}(\eta_{b})\rangle}{\langle v_{n}(\eta_{a})v_{n}(\eta_{b}) \rangle}.
\end{equation}

Similarly, decorrelation arising due to only event plane angle fluctuation is given by,

\begin{equation}
\label{eqn:3}
r_{n}^{\psi} = \frac{\langle \cos n(\Psi_{n}(-\eta_{a}) - \Psi_{n}(\eta_{b}))\rangle}{\langle \cos n(\Psi_{n}(\eta_{a}) - \Psi_{n}(\eta_{b}))\rangle}.
\end{equation}
\par
A recent study has proposed a novel observable called the four-particle cumulant ($T_{2}$), which remains unaffected by non-flow correlations. This observable proves to be sensitive to the type of flow plane decorrelation present, specifically distinguishing between S-shaped (torqued) decorrelation and C-shaped (bow) decorrelation~\cite{ref-t2}. In the S-shaped decorrelation, the flow plane angles in the forward and backward $\eta$ regions are located on opposite sides of the flow plane angle at mid-rapidity. In contrast, the C-shaped decorrelation occurs when the flow plane angles in the forward and backward $\eta$ regions fall on the same side as the flow plane angle at mid-rapidity. This new observable is given by,

\begin{equation}
\label{eqn:4}
T_{2} = \frac{\langle\langle \sin 2(\psi_{f} - \psi_{m,2}) \sin 2(\psi_{b} - \psi_{m,1})\rangle\rangle}{\langle \cos 2(\psi_{f} - \psi_{m,2})\rangle\langle \cos 2(\psi_{b} - \psi_{m,1})\rangle}
\end{equation}


Where the double bracket represents the cumulant and can be written in the form as shown in Eq.15. of Ref.~\cite{ref-t2}. $\Psi_{f}$ and $\Psi_{b}$ are the event plane angle in the forward and backward ($2 < |\eta_{(f/b)}| < 5$) $\eta$ window respectively. $\Psi_{m,1}$ and $\Psi_{m,2}$ are the two random sub-event plane angle in  $-1 < \eta <  0$ and $0 < \eta <  1$ region respectively.
In the absence of any specific decorrelation pattern, $T_{2}$ has a value of zero. A positive value indicates C-shaped decorrelation, while a negative value indicates S-shaped decorrelation.
\par
This paper is structured as follows: Sec.~\ref{sec2} provides a comprehensive description of the model employed in this study. In Sec.~\ref{sec3}, we present our results on the factorization ratios, $r_{2}$ and $r_{3}$ in Au+Au collisions at $\sqrt{s_{NN}}$ = 11.5 to 200 GeV, along with a thorough discussion. Finally, Sec.~\ref{sec4} offers a summary and concluding remarks on our study.

\section{Model description}
\label{sec2}
\label{sec2}
The AMPT model has four phases that comprehensively capture the entire evolution of a heavy-ion collision, from the initial stage to the late-stage hadronic interactions~\cite{ampt-ref}. The initial condition is obtained from a HIJING model~\cite{hijing-1, hijing-2, hijing-3} which includes phase-space distribution of minijet partons. The interactions among these partons are modeled by Zhang's parton cascade (ZPC)~\cite{zpc-1, zpc-2, zpc-3, zpc-4}. The differential cross section for parton-parton interaction is given by,

\begin{equation}
\label{eqn:a}
\frac{d\sigma}{dt} = \frac{9\pi\alpha^{2}}{2(t - \mu^{2})^{2}}.
\end{equation}

\begin{figure*}
\centering
\includegraphics[scale=0.6]{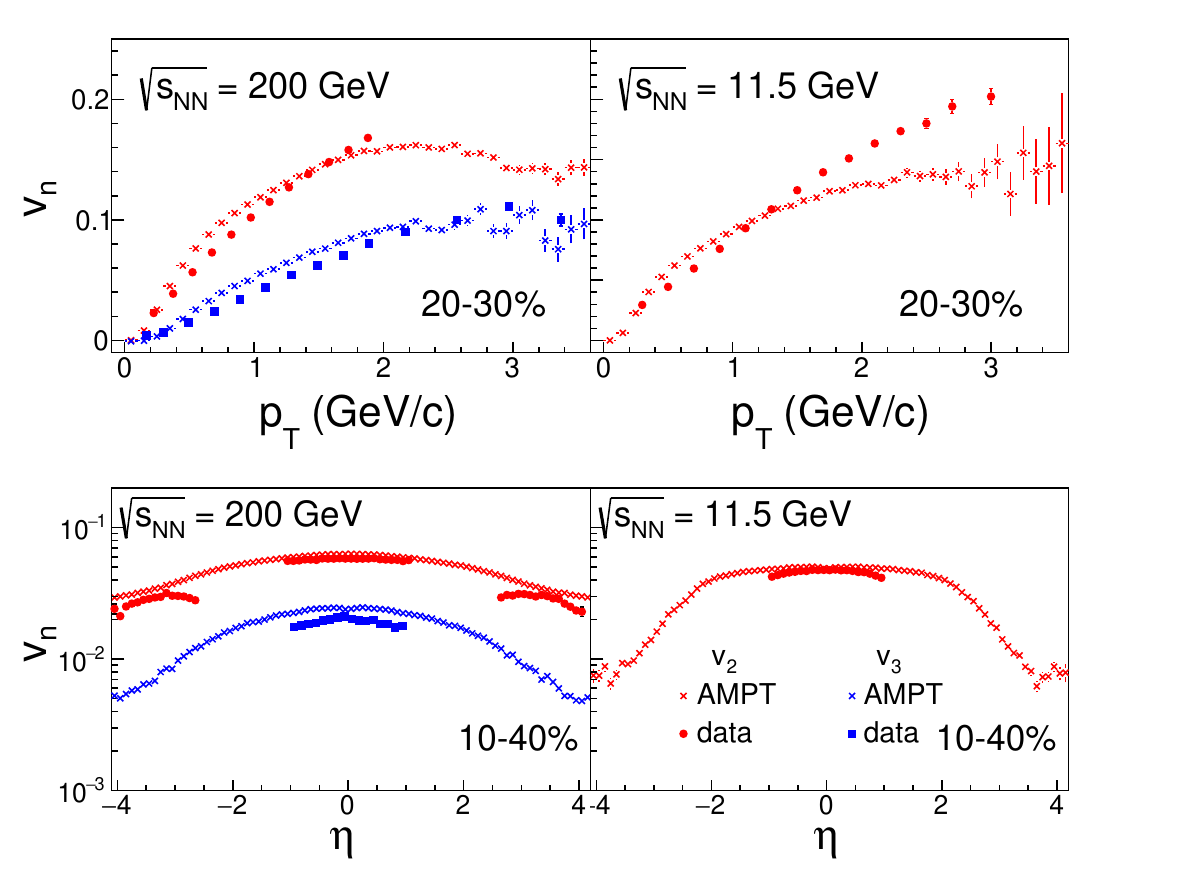}
\caption{Panel (a) and (b) show the transverse momentum dependent $v_{2}$ and $v_{3}$ in Au+Au collisions from AMPT model and its comparison with the data. Panel (c) and (d) show the pseudorapidty dependence of $v_{2}$ and $v_{3}$ compared with data. }
\label{fig:data}
\end{figure*}

Where $\sigma$ is the parton-parton scattering cross section, $t$ is the Mandelstam variable for four-momentum transfer, $\alpha$ is the coupling constant for strong interaction and $\mu$ is the Debye screening mass in partonic matter. For this study we have used the AMPT version v1.26t9b and  taken $\alpha$ = 0.33 , $\mu = $2.256 $fm^{-1}$ at all the energies, this will give us a parton-parton cross section of 3 mb. Taking this parameter set we have measured the transverse momentum and pseudorapidity dependence of elliptic flow ($v_{2}$) and triangular flow ($v_{3}$) at $\sqrt{s_{NN}}$ = 11.5 and 200 GeV as shown in Fig.~\ref{fig:data}. The measurement is compared with experimental data of RHIC ~\cite{star-data-v2-200,star-data-v2_eta-200, star-data-v3-200, star-v2-7-39} and we found a reasonable agreement between data and AMPT model.
We also generated the data with parton-parton cross section 10 mb to study the affect of parton scattering cross section on the observables $r_{2}$ and $r_{3}$.
Regarding the hadronization process, the AMPT model offers two distinct modes: the default mode and the string-melting mode. In the default mode, hadronization is implemented using the Lund string fragmentation model~\cite{lund-1, lund-2}. Conversely, in the string-melting mode~\cite{sm-1, sm-2, sm-3}, a quark coalescence model is employed. In this mode of hadronization, quarks which are close in space as well as in momentum will combine with each other to form hadrons. For our study, we have utilized the string-melting mode for hadronization.

\section{Results and Discussion}
\label{sec3}
\begin{figure*}
\centering
\includegraphics[scale=0.6]{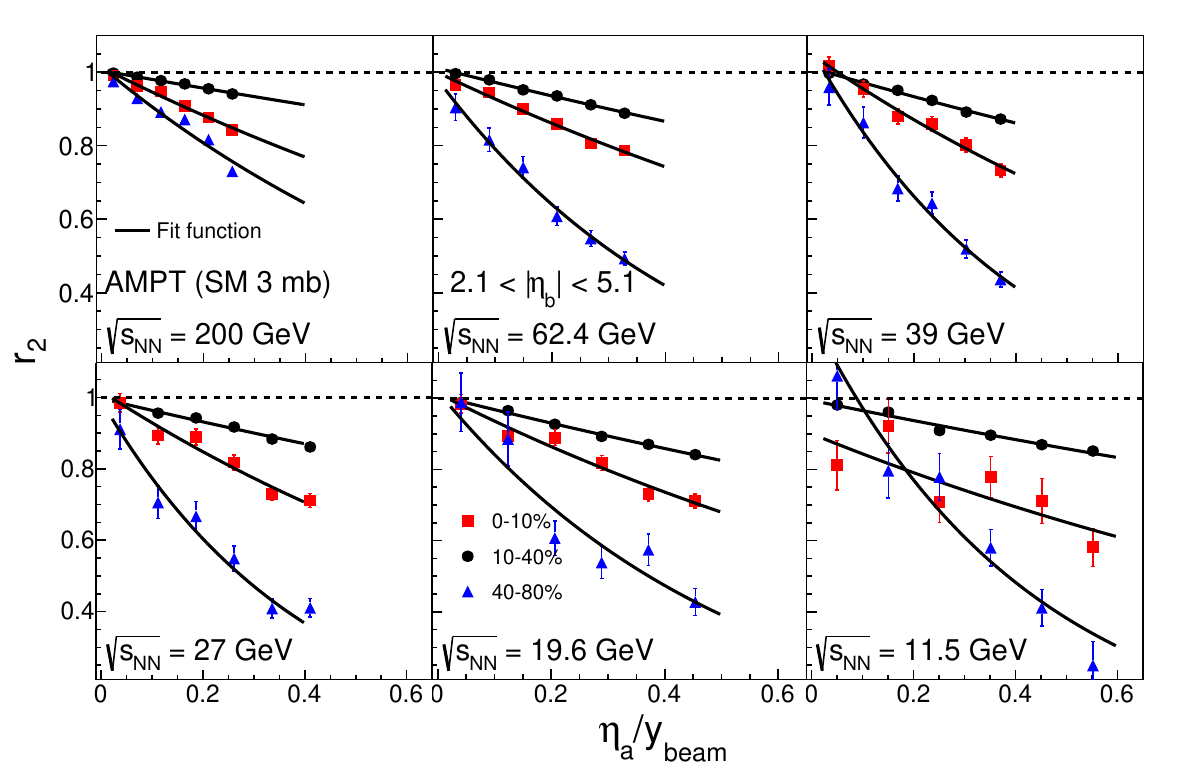}
\caption{$r_{2}$ is plotted as a function of $\eta_{a}/y_{beam}$ for Au+Au collisions in three centrality classes, 0-10\%,  10-40\%, and 40-80\% at six energies ranging from 11.5 to 200 GeV. The black solid line represents the fitting function: $e^{-2F_{2}\eta/y_{beam}}$. The vertical lines on the points represent the statistical uncertainty. }
\label{fig:r2}
\end{figure*}

\begin{figure*}
\centering
\includegraphics[scale=0.6]{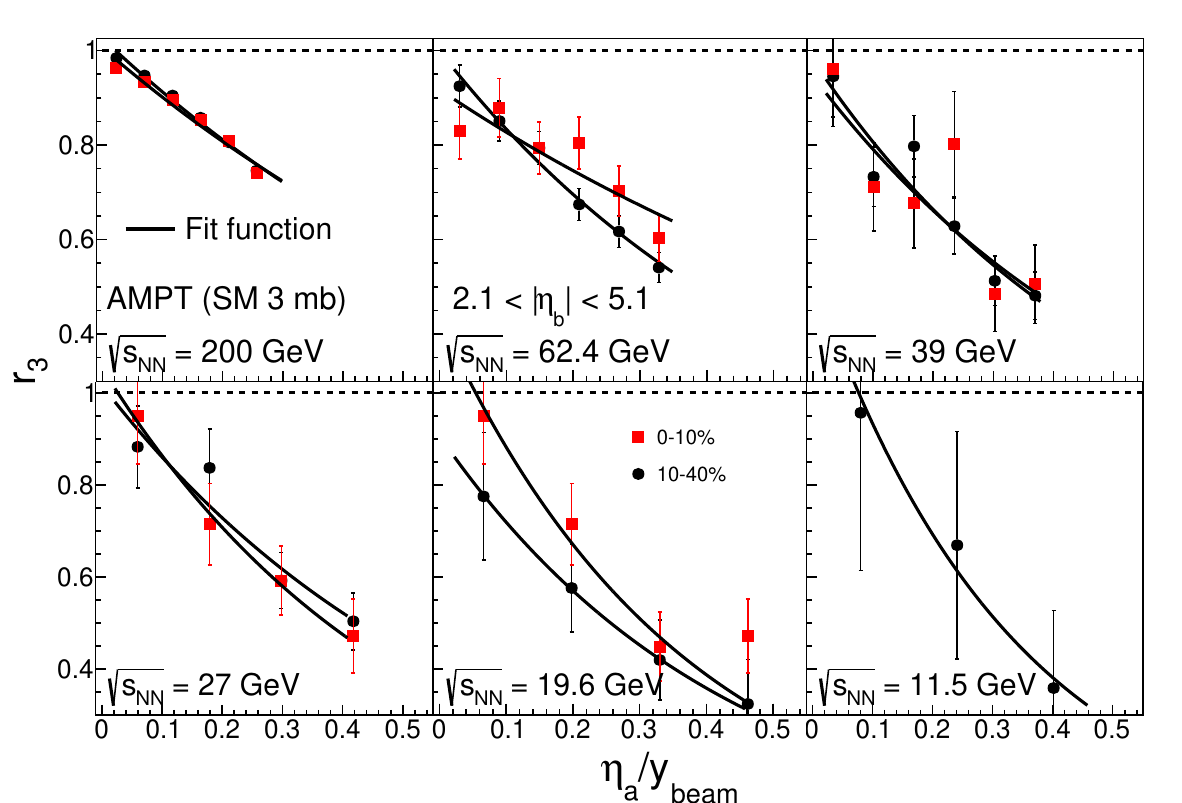}
\caption{$r_{3}$ is plotted as a function of $\eta_{a}/y_{beam}$ for Au+Au collisions in two centrality classes, 0-10\% and 10-40\% at six energies ranging from 11.5 to 200 GeV. The black solid line represents the fitting function: $e^{-2F_{3}\eta/y_{beam}}$. The vertical lines on the points represent the statistical uncertainty. }
\label{fig:r3}
\end{figure*}

In order to determine the values of $r_{2}$ and $r_{3}$ in Eq.~\ref{eqn:1}, specific windows are chosen for the variables $\eta_{a}$ and $\eta_{b}$. These windows are selected  keeping configuration of the STAR detector system in mind. For $\eta_{a}$, the window chosen is $-1.5 < \eta_{a} < 1.5$. This range corresponds to the acceptance of the Time Projection Chamber (TPC) detector in the STAR experiment~\cite{tpc-1, tpc-2}. For $\eta_{b}$, the window chosen is $2.1 < \eta_{b} < 5.1$. This range corresponds to the acceptance of the event plane detector (EPD) in the STAR experiment~\cite{epd}.

Figure~\ref{fig:r2} and Figure~\ref{fig:r3} show $r_{2}$ and $r_{3}$, respectively, as functions of $\eta_{a}/y_{beam}$ for Au+Au collisions at six different center-of-mass collision energies: $\sqrt{s_{NN}}$ = 11.5, 19.6, 27, 39, and 200 GeV measured over $p_{T}$ range 0.2 - 4.0 GeV/c in 0-10\%, 10-40\% and 40-80\% centrality events. Both $r_{2}$ and $r_{3}$ exhibit values below one and show a decreasing trend as $\eta_{a}/y_{beam}$ increases. This indicates a greater decorrelation with an increasing distance from the mid-rapidity region. A strong centrality dependence is observed in $r_{2}$. The magnitude of $r_{2}$ is found to be smallest in mid-central collisions wheres as in peripheral collisions the value of $r_{2}$ is the largest. 
This observation is consistent with the fact that in mid-central collisions the initial elliptic geometry dominate over the longitudinal fluctuation of the eccentricity as a result a smaller flow decorrelation is observed. In central and peripheral collisions, the elliptic flow is mainly driven by the fluctuations therefore we observe a strong decorrelation at these centralities. The centrality dependence of $r_{3}$ is weak because fluctuation is the dominant cause for triangular distribution of energy density. This trend remains consistent across all the colliding energies. 

\begin{figure*}
\centering
\includegraphics[scale=0.7]{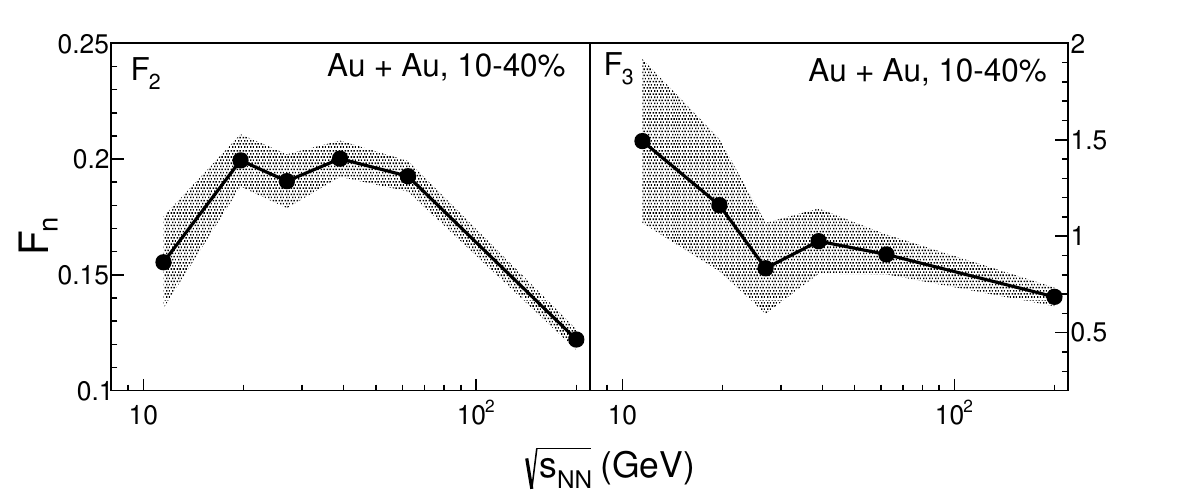}
\caption{Left panel shows $F_{2}$ as a function of $\sqrt{s_{NN}}$ in 10-40\% centrality events. The right panel shows the same for $F_{3}$. The shaded band represents the statistical uncertainty in the data.}
\label{fig:slope}
\end{figure*}

To quantify the energy dependence of the decorrelation, we fit $r_{2}$ and $r_{3}$ with a function $e^{-2F_{n}\eta_{a}/y_{beam}}$ as shown in Figures ~\ref{fig:r2} and \ref{fig:r3}. We extract the slope parameters, $F_{2}$ and $F_{3}$, and plot them as a function of the center-of-mass energy in Figure \ref{fig:slope}. We find that the value of $F_{3}$ is larger than the value of $F_{2}$ at all energies. In 10-40\% centrality, the average ratio $F_{3}/F_{2}$ is 5.2$\pm$0.3. 

As the collision energy decreases from $\sqrt{s_{NN}} = 200$ GeV to 62.4 GeV, both $F_{2}$ and $F_{3}$ exhibit an increasing trend. Below $\sqrt{s_{NN}} = 62.4$ GeV, $F_{2}$ shows an almost flat trend, while $F_{3}$ continues to rise consistently with further decreases in collision energy. Therefore $F_{3}$ found to be more sensitive to collision energy compared to $F_{2}$ in intermediate to lower RHIC energy regime. Such an increasing trend of decorrelation with decreasing collision energy indicates a stronger longitudinal decorrelation at lower collision energies.


In AMPT model, the effective shear viscosity to entropy density ($\eta_{s}/s$) is estimated by the following equation.
\begin{equation}
\label{eqn:6}
\eta_{s}/s \approx \frac{3\pi}{40\alpha^{2}}\frac{1}{(9+\frac{\mu^{2}}{T^{2}})\ln({\frac{18+\mu^{2}/T^{2}}{\mu^{2}/T^{2}}}) - 18}
\end{equation}
Where $\alpha$ is the coupling constant of strong interaction, $\mu$ is the Debye screening mass, and $T$ is the initial temperature of the heavy-ion collision estimated from the average energy density of mid-rapidity partons at their average formation time. The temperature $T$ is about 468 MeV at LHC and about 378 MeV at RHIC top energy~\cite{Junxu}. At $\sqrt{s_{NN}}$ = 19.6 GeV, the temperature of the QGP medium is taken to be 278 MeV. This assumption is motivated by the recent STAR measurement of dilepton production at finite baryon chemical potential~\cite{STAR-dilepton}. Therefore by varying $\alpha$ and $\mu$ we can change the $\eta_{s}/s$ and the parton-parton cross section of the medium given in Eq.~\ref{eqn:a}. 
We have varied $\alpha$ and $\mu$ as summarized in the Table ~\ref{table:1} to get different $\sigma_{pp}$ and $\eta_{s}/s$.  One can see that higher the cross section $\sigma_{pp}$ lower is the $\eta_{s}/s$. 

\begin{table}[h!]
\centering
\caption{Values of $\sigma_{pp}$ and $\eta_{s}/s$ for different values of $\alpha$ and $\mu$ in Au+Au collisions at $\sqrt{s_{NN}}$ = 200 and 19.6 GeV.}
\begin{tabular}{||c c c c c||} 
 \hline
 $\alpha$ & $\mu$ ($fm^{1}$) & $\sigma_{pp}$ (mb) & $\eta_{s}/s$ (200 GeV) & $\eta_{s}/s$ (19.6 GeV) \\ [0.3ex] 
 \hline\hline
  0.33 & 2.256 & 3 & 0.229 & 0.355 \\[1ex]
 0.47 & 1.8 & 10 & 0.086 & 0.126 \\[1ex]
 \hline
\end{tabular}
\label{table:1}
\end{table}

\begin{figure}[!htbp]
\centering
\includegraphics[scale=0.6]{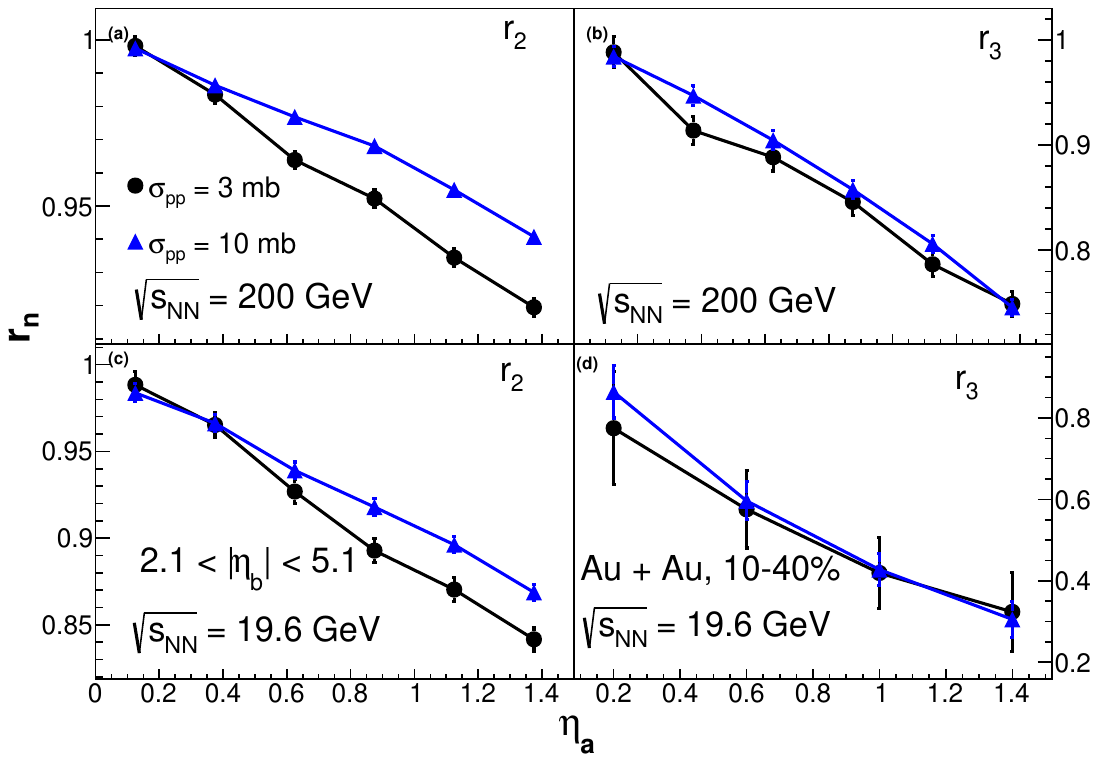}
\caption{Panel (a) and (b) show $r_{2}$ and $r_{3}$ as a function of $\eta$ at $\sqrt{s_{NN}}$ = 200 GeV. with parton-parton cross section 3 mb and 10 mb. Panel (c) and (d) show the same at $\sqrt{s_{NN}}$ = 19.6 GeV with parton-parton cross section 3 mb, and 10 mb.  }
\label{fig:sigma}
\end{figure}

The dependence of the parton-parton cross section on the observables $r_{2}$ and $r_{3}$ at $\sqrt{s_{NN}}$ = 19.6 and 200 GeV is illustrated in Fig.\ref{fig:sigma}. For a 3 mb cross section, $r_{2}$ is found to be greater compared to the 10 mb case. But $r_{3}$ is found to be independent of parton-parton cross section. This observation qualitatively aligns with previous findings reported in Ref.\cite{ref-model-1}, which were observed at LHC energy and continue to hold true at $\sqrt{s_{NN}}$ = 200 GeV and lower. Therefore studying flow decorrelation could be a potential observable to constrain the  $\eta_{s}/s$ of the medium.

\par
\begin{figure*}
\centering
\includegraphics[scale=0.6]{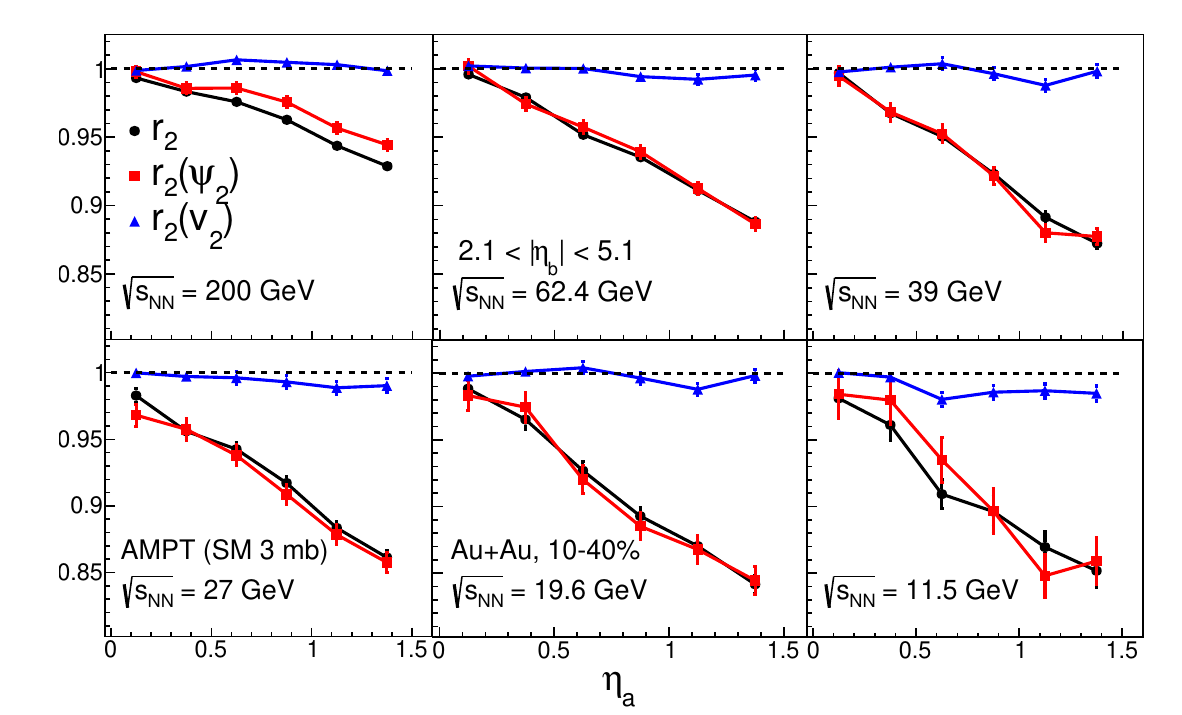}
\caption{The flow magnitude decorrelation ($r_{2}^{v}$), flow angle decorrelation ($r_{2}^{\psi}$) and the total decorrelation ($r_{2}$) is plotted as a function of $\eta_{a}$ for 10-40\% centrality in Au+Au collisions at 11.5 to 200 GeV.}
\label{fig:contribution}
\end{figure*}
\begin{figure*}
\centering
\includegraphics[scale=0.6]{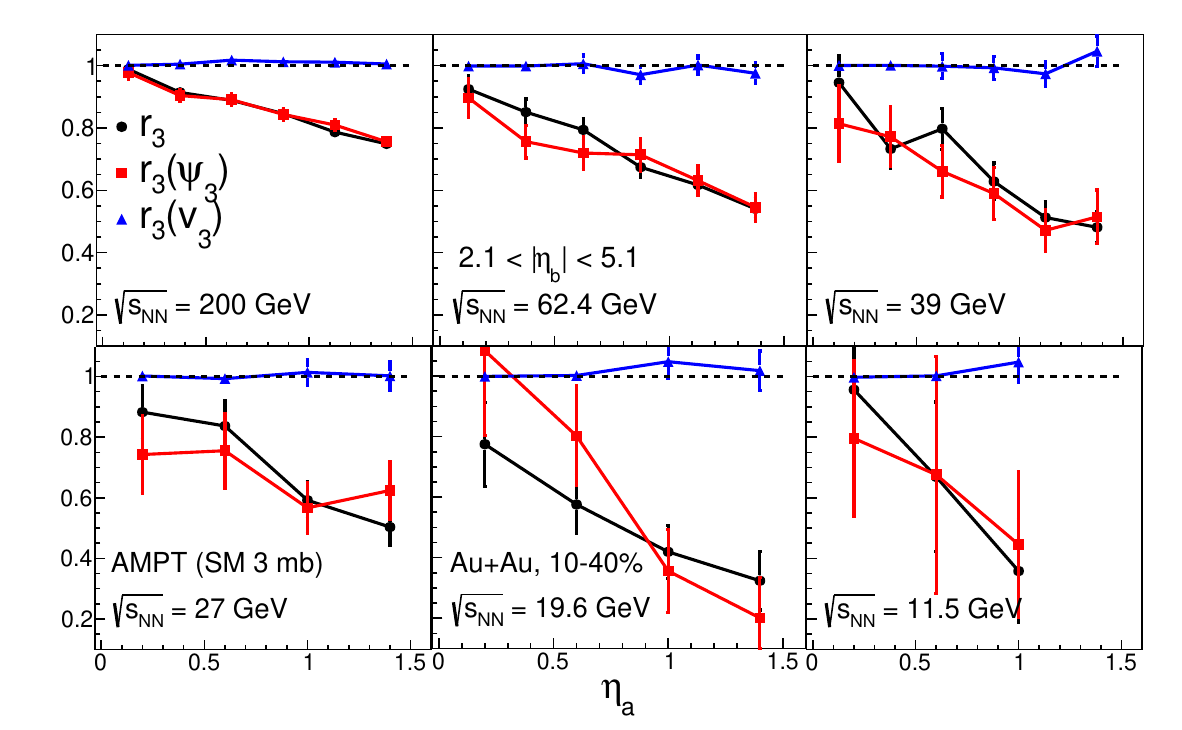}
\caption{The flow magnitude decorrelation ($r_{3}^{v}$), flow angle decorrelation ($r_{3}^{\psi}$) and the total decorrelation ($r_{3}$) is plotted as a function of $\eta_{a}$ for 10-40\% centrality in Au+Au collisions at 11.5 to 200 GeV.}
\label{fig:contribution_r3}
\end{figure*}

To understand the underlying factors contributing to the decorrelation, we examine the effects of flow magnitude and flow angle separately. The flow magnitude decorrelation is quantified using Equation~\ref{eqn:2}. Both $v_{n}(\eta_{a})$ and $v_{n}(-\eta_{a})$ are determined using the same event plane, extracted from the range $-0.5 < \eta < 0.5$. Similarly, the flow angle decorrelation is assessed using Equation~\ref{eqn:3}. Figure~\ref{fig:contribution} and Fig.~\ref{fig:contribution_r3} display the measured values of $r_{n}^{v}$ and $r_{n}^{\psi}$ alongside the total decorrelation $r_{n}$. The observation reveals that the primary contribution to the longitudinal flow decorrelation arises from the flow angle decorrelation, which is nearly equal to the total decorrelation $r_{n}$. On the other hand, the contribution from flow magnitude decorrelation is found to be negligible. This consistent pattern persists across all energies ranging from 11.5 to 200 GeV.

\par
\begin{figure}[!htbp]
\centering
\includegraphics[scale=0.5]{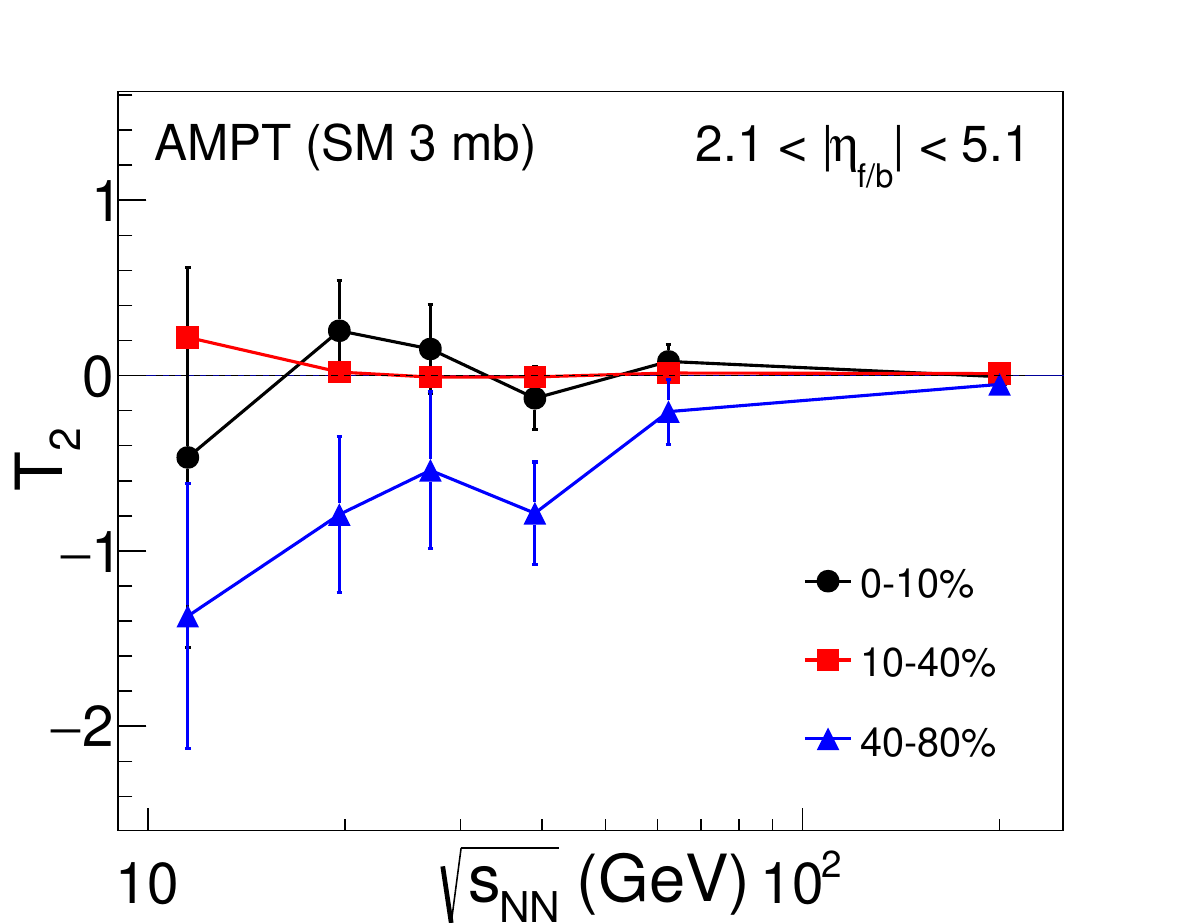}
\caption{The four-plane cumulant ($T_{2}$) is plotted as a function of center-of-mass energies in three different centralities, 0-10\%, 10-40\%, and 40-80\%. }
\label{fig:t2}
\end{figure}

The four-particle cumulant $T_{2}$ is measured as a function of center-of-mass energies in three different centralities with $2.1 < |\eta_{f(b)}| < 5.1$ as shown Figure~\ref{fig:t2}. For this study a transverse momentum cut, 0.15 $<p_{T}<$ 2.0 is used and only $\pi^{\pm}$,  $K^{\pm}$, and protons(anti-protons) are used for the analysis. In the central and mid-central collisions, 0-10\% and 10-40\%, the value of $T_{2}$ remains consistent with zero, suggesting a scenario where event plane angles calculated in each small $\eta$ interval are randomly oriented and do not depend on each other. In such a case we do not see any specific pattern of decorrelation.  
However for 40-80\% peripheral collisions, we observe a consistent hint of negative $T_{2}$ in all the energies which suggests the presence of S-shaped decorrelation in peripheral collisions. The S-shaped decorrelation in 40-80\% is consistent with the idea of a torqued energy deposition along rapidity direction in peripheral collisions. The magnitude of $T_{2}$ in peripheral collisions are found to larger in smaller collision energies indicating a strong decorrelation scenario in lower energy regimes. 
\par

\section{Summary}
\label{sec4}
In summary, we have presented the measurement of longitudinal flow decorrelation parameters, $r_{2}$ and $r_{3}$, in Au+Au collisions at RHIC BES energies, ranging from $\sqrt{s_{NN}}$ = 11.5 to 200 GeV using the AMPT model.  We observe that $r_{3}$ is more sensitive the collision energy compared to $r_{2}$. 


Furthermore, we present the measurement of the four-plane cumulant $T_{2}$. Our analysis provides a hint of negative value of $T_{2}$ at all energies in 40-80\% collision centralities indicating an S-shaped flow plane decorrelation but the present result is not significant considering the large statistical fluctuation in 40-80\% central events. Whereas in 0-10\% and 10-40\% centrality, $T_{2}$ is consistent with zero, this indicates the absence of any specific pattern of decorrelation where event plane angles are oriented randomly at each $\eta$ intervals.
\par
Measurement of both $r_{2}$ and $r_{3}$ in RHIC BES energies could be potential observables to constrain $\eta_{s}/s$ of the QGP medium and shed light on the collision energy dependence of $\eta_{s}/s$.


\begin{thebibliography}{100}
\footnotesize
\small


\bibitem{QGP} E. Shuryak,  Rev. Mod. Phys. 89, 035001 (2017).
\bibitem{QGP-2}W. Busza, K. Rajagopal, and W. van der Schee, Ann. Rev. Nucl. Part. Sci. 68, 339 (2018).
\bibitem{flow1} J. Y. Ollitrault,  Phys. Rev. D 46, 229 (1992).
\bibitem{flow2} P. Huovinen, P. F. Kolb, U. Heinz, P. V. Ruuskanen, and S.A. Voloshin {\it et al.}, Phys. Lett. B 503, 58 (2001).
\bibitem{flow3} C. Shen and U. Heinz, Phys. Rev. C 85, 054902 (2012).
\bibitem{flow4} R. Snellings, New J. Phys. 13, 055008 (2011).
\bibitem{flow5}H. Appelshauser {\it et al.}, (NA49 Collaboration), Phys. Rev. Lett. 80, 4136 (1998).
\bibitem{flow6} K. Aamodt {\it et al.}, (ALICE Collaboration), Phys. Rev. Lett. 105, 252302 (2010).
\bibitem{flow7} G. Aad {\it et al.}, (ATLAS Collaboration), Phys. Lett. B 707, 330 (2012).
\bibitem{flow8} S. Chatrchyan {\it et al.} (CMS Collaboration), Phys. Rev. C 89, 044906 (2014)
\bibitem{exp-1}V. Khachatryan {\it et al.} (CMS Collaboration), Phys. Rev.C 92, 034911 (2015).
\bibitem{exp-2}M. Aaboud {\it et al.} (ATLAS Collaboration), Eur. Phys. J.C 78, 142 (2018).
\bibitem{exp-3}G. Aad et al. (ATLAS Collaboration), Phys. Rev. Lett. 126, 122301 (2021).
\bibitem{exp-4}M. Nie (STAR Collaboration), Nucl. Phys. A 1005, 121783 (2021).
\bibitem{ref-model-1}L. Pang, G. Qin, V. Roy, X. Wang, and G. Ma, Phys. Rev. C 91, 044904 (2015).
\bibitem{ref-model-2}L. Pang, H. Petersen, G. Qin, V. Roy, and X. Wang, Eur. Phys. J. A 52 (2016) 4, 97.
\bibitem{ref-model-3} W. Zhao, S. Ryu, C. Shen, and B. Schenke, Phys. Rev. C 107, 014904 (2023).
\bibitem{ref-model-4}P. Dasgupta, H. Wang, G. Ma, Phys. Rev. C 107, 014905 (2023).
\bibitem{ref-model-5}A. Sakai, K. Murase, T. Hirano, Phys. Lett. B 829, 137053 (2022).
\bibitem{ref-model-6}A. Sakai, K. Murase, T. Hirano, Phys. Rev. C 102, 064903 (2020).
\bibitem{ref-model-7}A. Behera, M. Nie, J. Jia, Phys. Rev. Res. 2, 023362 (2020).
\bibitem{ref-model-8}X. Wu, L. Pang, G. Qin, X. Wang Phys. Rev. C 98, 024913 (2018).
\bibitem{ref-model-9}P.Bozek, W. Broniowski, Phys. Lett. B 752, 206-211 (2016).




\bibitem{ref-1}M. Aaboud et al. (ATLAS Collaboration), Eur. Phys. J.C 78, 142 (2018).
\bibitem{ref-2}P. Bozek and W. Broniowski, Phys. Rev. C 97, 034913 (2018).
\bibitem{ref-3}J. Cimerman, I. Karpenko, B. Tomášik, and B. A. Trzeciak, Phys. Rev. C 104, 014904 (2021).
\bibitem{ref-t2}Z. Xu, X. Wu, C. Sword, G. Wang, S. Voloshin, and H.Z. Huang, Phys. Rev. C 105, 024902 (2022)
\bibitem{ampt-ref}Z.W.Lin, C.M.Ko, B.A.Li, B.Zhang and S.Pal, Phys. Rev. C 72, 064901 (2005).
\bibitem{hijing-1}X. N. Wang, Phys. Rev. D 43, 104 (1991).
\bibitem{hijing-2}X. N. Wang and M. Gyulassy, Phys. Rev. D 44, 3501 (1991).
\bibitem{hijing-3}X. N. Wang and M. Gyulassy, Phys. Rev. D 45, 844 (1992).
\bibitem{zpc-1}G. C. Rossi and G. Veneziano, Nucl. Phys. B123, 507 (1977).
\bibitem{zpc-2}L. Montanet, G. C. Rossi, and G. Veneziano, Phys. Rep. 63, 149 (1980).
\bibitem{zpc-3}D. Kharzeev, Phys. Lett. B 378, 238 (1996).
\bibitem{zpc-4}V. ToporPop, M. Gyulassy, J. Barrette, C. Gale, X. N. Wang, and N. Xu, Phys. Rev. C 70, 064906 (2004).
\bibitem{star-data-v2-200} J. Adams {\it et al}. (STAR Collaboration) Phys. Rev. C 72, 014904 (2005).
\bibitem{star-data-v2_eta-200} B. I. Abelev {\it et al}. (STAR Collaboration) Phys. Rev. C 77, 054901 (2008).
\bibitem{star-data-v3-200}L. Adamczyk{\it et al}. (STAR Collaboration) Phys. Rev. C 88, 014904 (2013).
\bibitem{star-v2-7-39}L. Adamczyk{\it et al}. (STAR Collaboration) Phys. Rev. C 86, 054908 (2012).
\bibitem{lund-1}B. Andersson, G. Gustafson, and B. Soderberg, Z. Phys. C 20, 317 (1983).
\bibitem{lund-2}B. Andersson, G. Gustafson, G. Ingelman, and T. Sjostrand, Phys. Rep. 97, 31 (1983).
\bibitem{sm-1}Z. W. Lin and C. M. Ko, Phys. Rev. C 65, 034904 (2002).
\bibitem{sm-2}Z. W. Lin, C. M. Ko, and S. Pal, Phys. Rev. Lett. 89, 152301 (2002).
\bibitem{sm-3}Z. W. Lin and C. M. Ko, J. Phys. G 30, S263 (2004).
\bibitem{tpc-1} K.H. Ackermann {\it et al}. (STAR Collaboration) Nuclear Instruments and Methods in Physics Research A 499 624–632 (2003).
\bibitem{tpc-2} C. Yang (STAR collaboration) Nucl. Phys. A 967 800–803 (2017).
\bibitem{epd} J. Adams {\it et al.} (STAR collaboration) Nuclear Instruments and Methods in Physics Research A 968, 163970 (2020).
\bibitem{Junxu} J. Xu and C. M. Ko Phys. Rev. C 83, 034904 (2011).
\bibitem{STAR-dilepton} (STAR Collaboration) arXiv:2402.01998.










 


\end{thebibliography}
\end{document}